\documentclass[runningheads]{llncs}
\usepackage{graphicx}
\usepackage{url}

\begin{document}
\title{SketchMeHow: Interactive Projection Guided Task Instruction with User Sketches}


\author{Haoran Xie \and
Yichen Peng \and
Hange Wang \and
Kazunori Miyata}
\authorrunning{H. Xie et al.}
\titlerunning{SketchMeHow}
%
\institute{Japan Advanced Institute of Science and Technology, Ishikawa 9231292, JAPAN \\
\email{xie@jaist.ac.jp}}
\maketitle             
\begin{abstract}
In this work, we propose an interactive general instruction framework SketchMeHow to guidance the common users to complete the daily tasks in real-time. In contrast to the conventional augmented reality-based instruction systems, the proposed framework utilizes the user sketches as system inputs to acquire the users' production intentions from the drawing interfaces. Given the user sketches, the designated task instruction can be analyzed based on the subtask division and spatial localization for each task. The projector-camera system is adopted in the projection guidance to the end-users with the spatial augmented reality technology. To verify the proposed framework, we conducted two case studies of domino arrangement and bento production. From our user studies, the proposed systems can help novice users complete the tasks efficiently with user satisfaction. We believe the proposed SketchMeHow can broaden the research topics in sketch-based real-world applications in human-computer interaction\footnote{This is a pre-peer review, pre-print version of this article. The final authenticated version is available online at proceedings of HCII 2021(International Conference on Human-Computer Interaction).}.   

\end{abstract}
\section{Introduction}

In our daily lives, numerous tasks required special skills or adequate experience, which usually frustrate the common and amateur users to make any attempt. In addition, the daily tasks may be tedious and complex procedures, and it is difficult and challenging to conduct creative activities. In the research fields of creativity support in human-computer interaction, the previous works mainly focused on the graphical user interface in the computer environment. In this work, we aim to help the common user to complete the designated tasks under the computed guidance in real-world applications. 

To solve this issue, the common approaches are spatial computing techniques of instruction systems in a workspace, such as augmented reality approaches with head-mounted display and projection, see-through displays such as tablet \cite{instruction2018,instruction2021}. Considering the system stability and cognitive loads of the instruction systems, we adopt the spatial augmented reality technology with projector-camera system \cite{instruction2016}. Especially, the depth camera was used for sensing the depth information for instruction generation rather than the standard web cameras. In the previous instruction works \cite{instruction2016,smith20}, the users are asked to complete the designated task with the given target where the requirement of satisfying the users’ design intention is absent. In this work, we combine the creativity support and instruction systems to help users to conduct tasks with their own designs. To achieve this research purpose, we utilized the freehand user sketches as system input which are difficult to be handled due to the ambiguous and abstract visual representations. 

\begin{figure}[t]
\includegraphics[width=\textwidth]{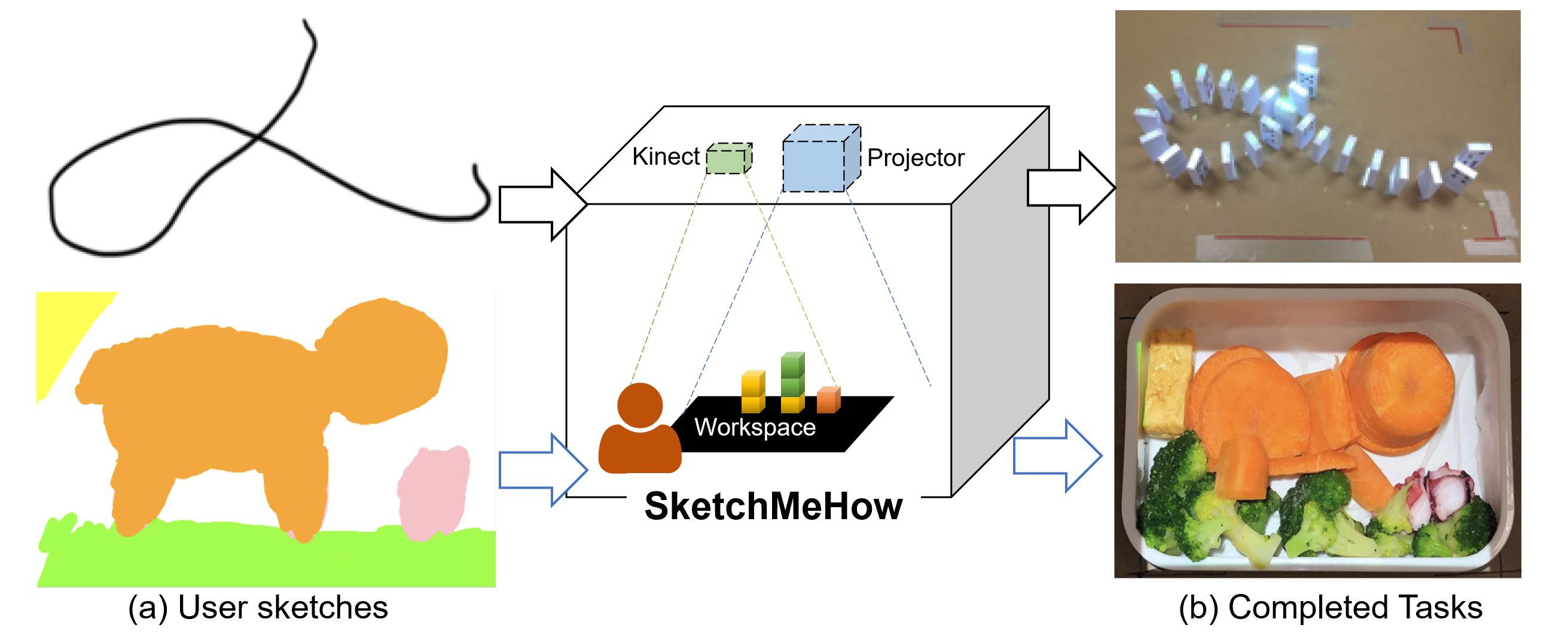}
\caption{The proposed interaction framework of SketchMeHow.} 
\label{fig:concept}
\end{figure}

Therefore, we propose an interactive instruction system, SketchMeHow that can provide the users with projection-based guidance in real-time with spatial augmented reality technique. As shown in Figure \ref{fig:concept}, the system input of SketchMeHow may include freehand line drawing or color illustration. With the help of user guidance in projector-camera systems, the completed tasks can be achieved at low time costs and high success rates.

To verify the proposed SketchMeHow framework, we conducted two usage applications for task instruction from user sketches: domino block arrangement and bento production. In the block arrangement system, the stroke input is analyzed with the individual block, and the gradient colors are projected on the blocks for guiding the correct positions. Finally, the domino blocks can be collapsed successfully. In the bento production system, we adopt both stroke and color of user sketches as input. The food ingredients are recommended based on user input and arranged with the projected color guidance. The accomplished bento is in compliance with the user sketches well. From our user study and subjective questionnaires, we verified that the proposed framework of SketchMeHow can provide the interactive projection guidance to help non-skilled users to complete theirs desired target in specific tasks. The proposed systems are effective and useful compared with the conventional approaches without guidance. We believe SketchMeHow can facilitate sketch-based systems for real-world applications. 

\section{Related Work}

\subsection{Sketch-based Interface}
Freehand sketches can present the users' design intention and have been adopted intensively in two-dimensional content design. Teddy proposed the sketch-based interface to construct 3D polygonal models from freehand drawn sketches \cite{IgarashiMT99}. A sketch-based interface was usually adopted in the image and shape retrieval \cite{EitzRBHA12} and drawing guidance \cite{XieHLW14}. Recently, the deep learning algorithms were widely developed in the sketch user interface, such as facial image editing \cite{FaceShop18} and drawing \cite{huang21dualface}, anime image editing \cite{sketch2anime}, image translation \cite{ghosh2019interactive} and normal map estimation \cite{sketch2normal}. Besides static applications of image editing, the physical dynamics were controlled with user sketches such as hair \cite{brush2016} and fluid \cite{zhu19}. In this work, we especially focus on the sketch interfaces for task instructions with augmented reality approaches. 

\subsection{Sketch-based Guidance}
 It is challenging and useful to adopt the sketch-based design for real-world applications with interactive user guidance. Sketch and run used the sketch-based interface for robot control based on a top-down camera view \cite{sketchrun2009}. Lighty proposed the guidance system of room lights by sketching a target illumination area on tablet devices \cite{noh13}. However, these proposed systems were designed for automatic machine control where the user guidance was absent such task instruction. To solve this issue, Sketch2Domino proposed a sketch-based guidance system for domino toppling production \cite{peng2020sketch2domino}. Sketch2Bento proposed the guide system for ingredient arrangement of designing lunch boxes \cite{sketch2bento}. In this work, we aim to propose a general instruction framework for sketch-based user guidance in real-world applications.

\subsection{Projection-based Guidance}
The spatial augmented reality adopts interactive projection for various real-world applications. Illuminating clay was proposed as a tangible interface for landscape design \cite{mit2002}, Flagg et al. worked on the implementation to guide oil painting using the projection techniques \cite{flagg2006projector}.  There are several works to adopt projection-based guidance in fabrication purposes, such as sculpting \cite{river2012}. For large-scale fabrication, BalloonFab guided the user to create the large-scale balloon art \cite{xie2019balloonfab}. A similar approach was used in newspaper sculpture \cite{newsfab}. For daily activities, the projection guidance can also be helpful in solving the puzzle of Rubik's cubes \cite{ajisaka2020learning}, calligraphy practice \cite{he2020}, and sports training \cite{sano2016sports}. For room-scale projection, Roomalive was proposed for multiple projector-camera systems \cite{roomalive}.  In contrast to these projection-based guidance approaches, we first adapt user sketches as system control to meet user design intentions.

\section{Framework Overview}
\begin{figure}[t]
\includegraphics[width=\textwidth]{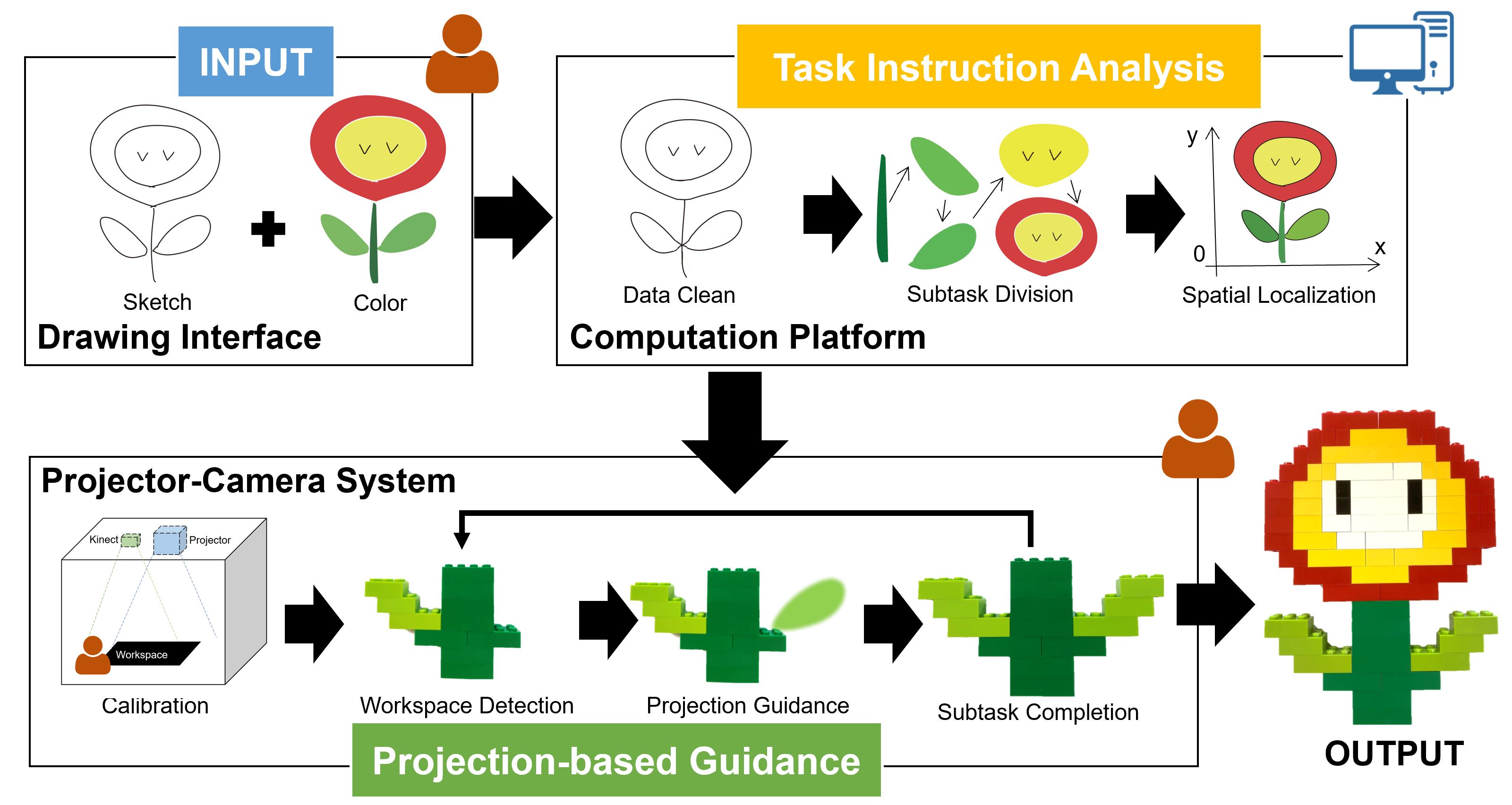}
\caption{The proposed interaction framework of SketchMeHow.} 
\label{fig:framework}
\end{figure}

Figure \ref{fig:framework} shows the proposed framework of SketchMeHow which is composed of three modules on the separated platforms: system input, task instruction analysis, and projection-based guidance. For the input module, a drawing user interface is developed with the stoke and coloring function with pen-based interaction on a tablet or real workspace. For the task instruction analysis module, the input hand-drawn sketches are cleaned with stroke smoothing to exclude the infeasible instructions. According to the requirements of specific works, we divide the whole sketch instruction into separated sub-tasks with the elemental objects, such as individual blocks in element construction. Finally, the locations of each task object are calculated based on the subtasks. For the projection-based guidance module, we adopt the projector-camera system with the depth camera. The system calibration is conducted based on multi-pin correction to adjust the projector and camera coordinates. Then, the current workspace is detected with depth differences from the captured depth maps. With the calculated task instruction, the guidance information is projected onto the workspace overlapped with the calculated operations. After the iterative processes of subtasks, the desired productions can be achieved with guidance. 

\subsection{Drawing Interface}
\begin{figure}[t]
\centering
\includegraphics[width=1.0\linewidth]{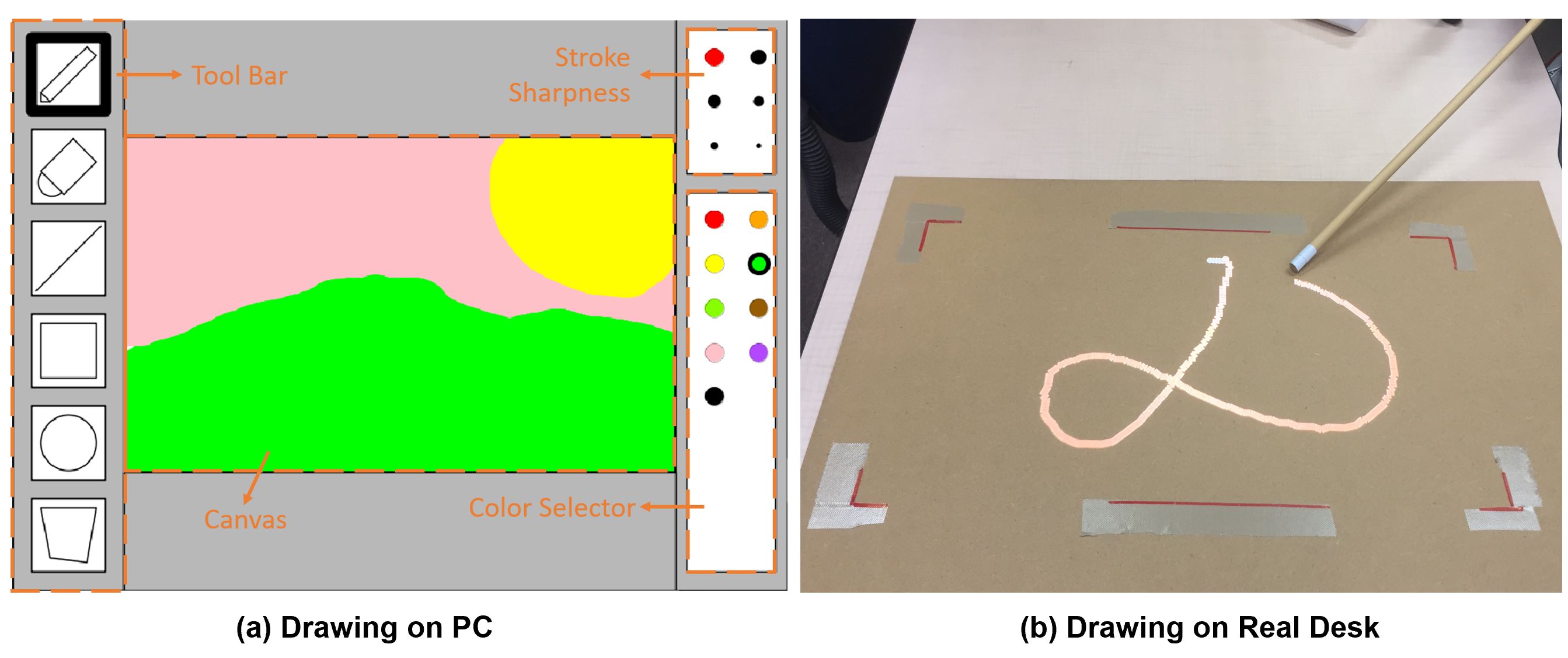}
\caption{Drawing interface can be implemented as graphical user interface on computer (a) or the 3D spatial interface on real desk (b).}
\label{fig:drawing}
\end{figure}

The sketching interface can be developed based on the graphical user interface or tangible physical interface. The examples of user interfaces are shown in Figure \ref{fig:drawing}. For the graphical drawing interface, the canvas is used for stroke-based interaction as Figure \ref{fig:drawing}(a). Besides the canvas, the user can select tools from the toolbars includes pencil, eraser, line, and different shapes. In addition, the user can select the stroke sharpness and stroke color. For the physical drawing interface, it is convenient for the user to draw with a pen-type input device. A simple solution is to use the retro-reflective marker on the pen tip in a depth camera-projector system as Figure Figure \ref{fig:drawing}(b). In our case studies, we implemented both graphical and physical drawing interfaces in the SketchMeHow framework.  

\subsection{Spatial Augmented Reality}
\begin{figure}[htb]
\centering
\includegraphics[width=0.9\linewidth]{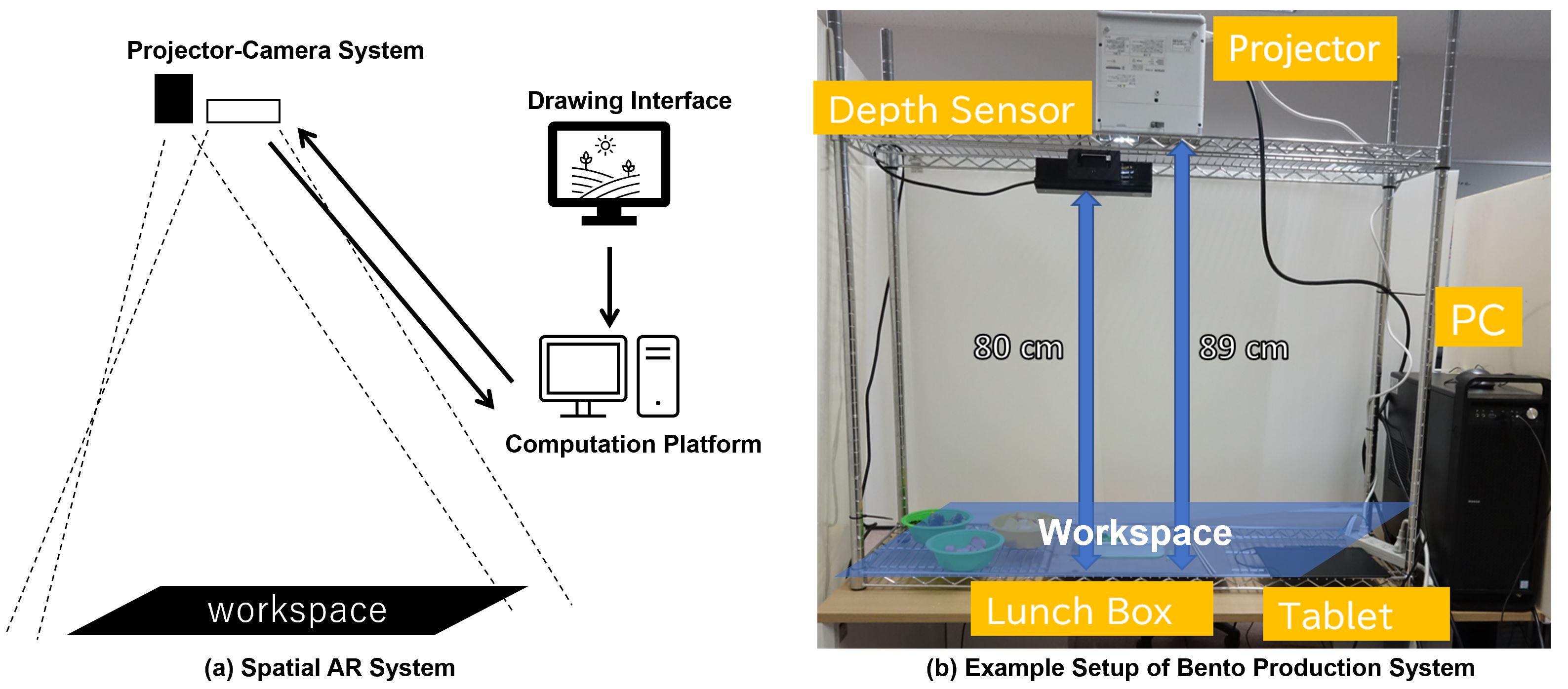}
\caption{Spatial augmented reality system (a); an example setup for bento design (b).}
\label{fig:setup}
\end{figure}

Due to the absence of head-mounted devices, it is more convenient in instruction guidance systems using interactive projection \cite{instruction2016}. As shown in Figure \ref{fig:setup}(a), a projector-Camera System is usually adopted in the setup of spatial augmented reality system. The analyzed results from the drawing interface are projected on the workspace. In the prototype systems in our case study, we used a depth camera to capture the depth map or image recognition in infrared frames. An example setup of the proposed system is shown in \ref{fig:setup}(b). We used the pin-points tool to ensure the appropriate calibration of projector-camera system. 

\begin{figure}[htb]
\centering
\includegraphics[width=0.9\linewidth]{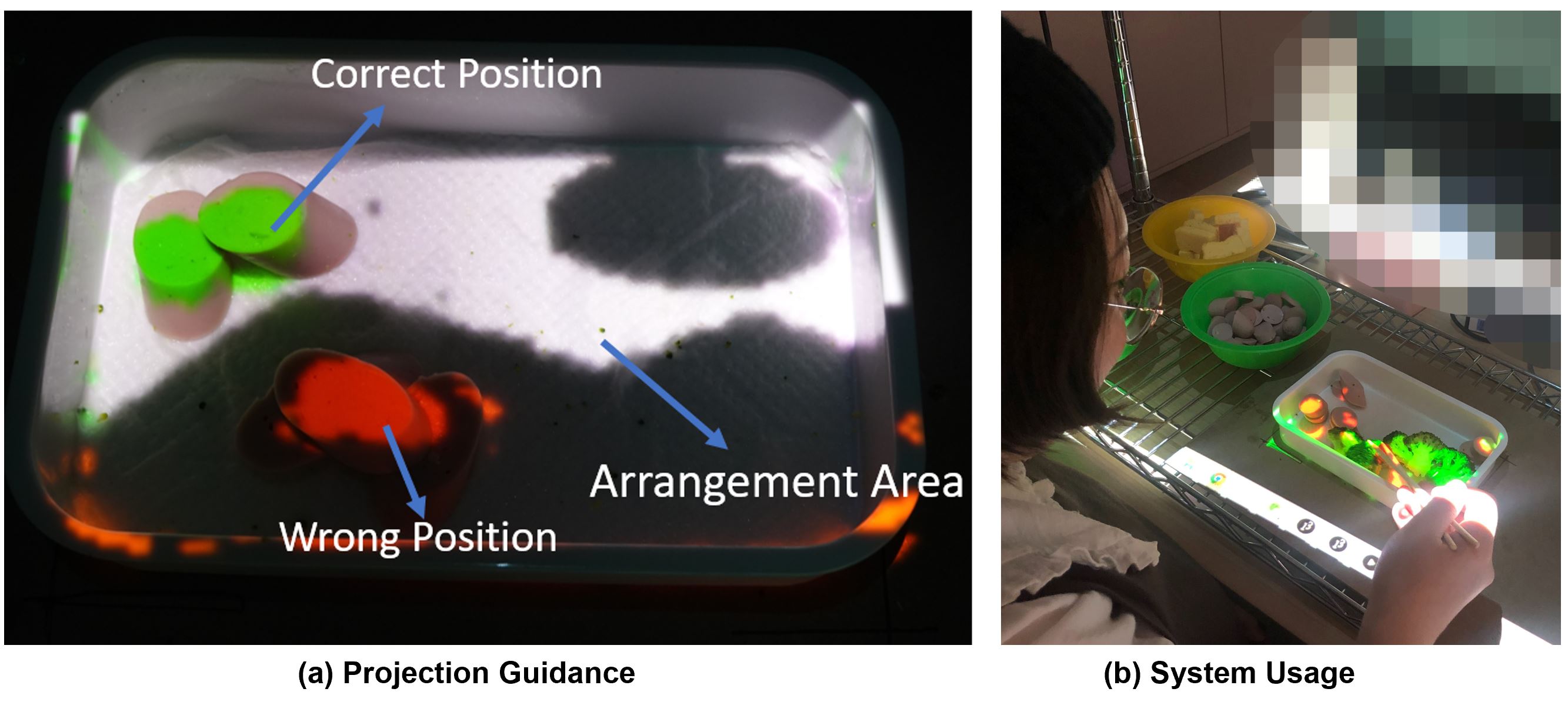}
\caption{Projection guidance for task instruction (a) and the system usage of the proposed SketchMeHow framework with bento production (b).}
\label{fig:bentoguidance}
\end{figure}

\subsection{User Guidance}
In our prototype systems, the depth data of captured depth map is used to guide the arrangement of instruction elements. For the projector calibration, the guidance image projected onto the workspace is distorted to match the coordinates of the depth view. In addition, we stored the pixel positions for each color in the scaled guidance image to provide guidance for each sub-tasks. The guidance image is projected onto the workspace with different colors to help complete tasks as shown in Figure \ref{fig:bentoguidance}(a). We map the stored color list to the depth data to obtain the areas, and the user can place the objects under the corresponding tasks. For example, a black color denotes the area where the object should not be placed. In addition, the green color refers to the correct placement, and the red color for incorrect placement. An example of the usage of the proposed instruction system is shown in Figure \ref{fig:bentoguidance}(b).

\section{Case Study}
To verify the proposed SketchMeHow framework, we conducted two case studies of domino arrangement and bento production.

\subsection{Domino Arrangement}
To support designing a complex chain reaction such as domino blocks, we implemented a sketch-based instruction system for the domino placement according to the user customized sketch in this case study \cite{peng2020sketch2domino}. The prototype system applied interactive projection-mapping technology to guide users to places the domino blocks. Users were allowed to design the blueprint of the arrangement by the freehand sketches without considering the physical limitations of domino placement following the rules of chain reaction.

\subsubsection{System Overview}
\begin{figure}[t]
    \centering
    \includegraphics[width=1.0\linewidth]{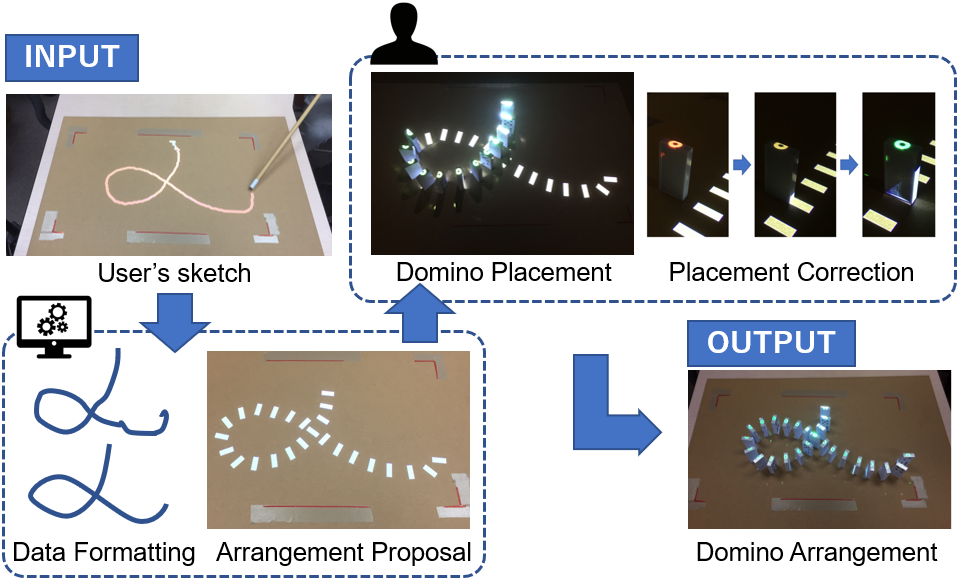}
    \caption{The prototype system of the proposed domino arrangement system.}
    \label{fig:dominosystem}
\end{figure}

As shown in Figure \ref{fig:dominosystem},  users can draw the sketch for the intention design of the domino arrangement. The proposed system analyzed the stroke data of input sketches to make the domino blocks collapse successfully in a valid arrangement where the vertices are excessively dense or the angles are quite acute. After that, the users can locate the domino blocks following the projection guidance to confirm the target positions. 

\subsubsection{System Environment}
Similar to the example setup in Figure \ref{fig:setup}, we set the working space (70 cm-high desk) with a laser projector (LG HF80JG, 2000 lumen), and a depth camera (Microsoft Kinect V2, the resolution of depth image is 512 x 424). The projector and camera were set vertically 85.1 cm and 77.8 cm above the desk. The size of the projection area was 57.2 cm x 32.1 cm. In this study, standard domino blocks (23 mm wide,  46 mm high and 8 mm thick) were used which was certified by the Japan Domino Association. 

\subsubsection{Projection Guidance}
The users can use a pen stick with the marker to draw a sketch directly on the projection area. The arrangement proposal according to the drawn target is projected as visual guidance of the valid domino's positions, so that the user can arrange the domino blocks in the order following the design.  For the correction of the valid domino's position, a gradient-colored circle is projected at the center of the placed domino blocks for the user to confirm whether they have been placed correctly. The colors of the circles are set in gradient colors in terms of the distance from the target position. The circle on the top of the block is in red color if the distance is larger than 2.0 cm from the target position. The circle turns yellow about 1 cm away. The circle turns green when is correctly placed. Note that the target positions of domino are shown by white rectangles from the beginning.

\subsubsection{User Study} 
We conducted a comparison study to verify the effectiveness of the proposed system. we compared the domino arrangement with and without the proposed system. In conventional approach without the systems, the participants were asked to place the domino blocks following the hand-drawn sketches on a printed paper without explicit guidance on the intervals between the domino blocks. The criterion for a successful design was that all the domino blocks could collapsed with one simple push. 12 graduate student participants (11 males, 1 female) joined the comparison study.  

 For our prototype system, the participants were asked to arrange the domino blocks with the help of the visual guidance. The user can start the arrangement from any location, because the arrangement order is non-trivial. Note that we adopted the same blueprint design used in the conventional approach. Both operational time costs and the results were recorded. In addition, we conducted a questionnaire for the subjective evaluations of the prototype system. The questionnaire had four questions: a) Which approach is easier for arranging domino blocks? b) Which perfection do you have confidence in? c) Which result gave you a sense of accomplishment? d) Which approach was more interesting?

\begin{figure}[t]
    \centering
    \includegraphics[width=1.0\linewidth]{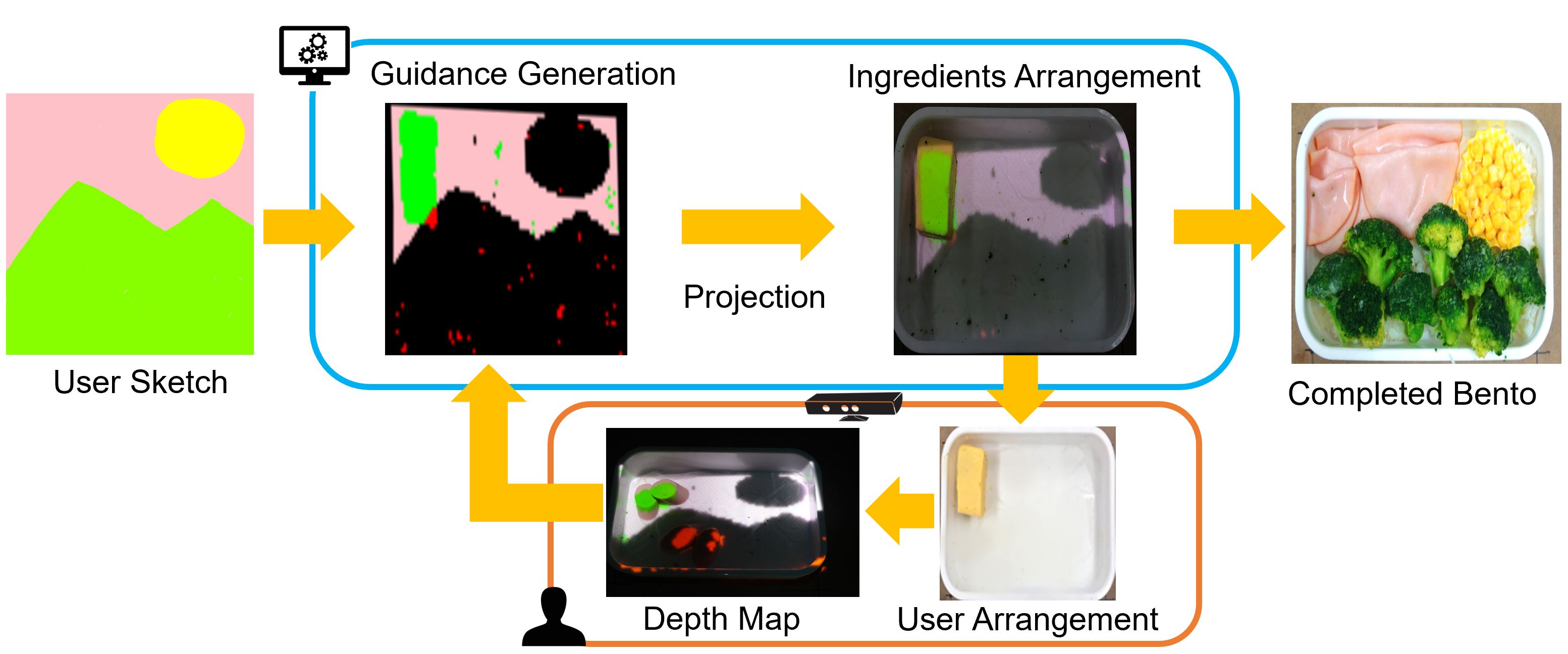}
    \caption{The prototype system of the proposed bento production system.}
    \label{fig:bentosystem}
\end{figure}

\subsection{Bento Production}
In this case study, we implemented an ingredient placement support system for bento (lunch box) production based on the user hand-drawn sketches. The prototype system allows users to make the bento following creative intentions without relying on existing recipes.

\subsubsection{System Overview} As illustrated in Figure \ref{fig:bentosystem}, the user first sketches the desired arrangement image with the drawing interface. Then, the system provides the projection guidance to lunch box from the sketch input. Under the help of the production guidance, the user can complete the ingredient arrangement for desired bento. Finally, the real-time guidance is achieved based on the current working progress.

\subsubsection{System Environment}
As shown in Figure \ref{fig:setup}, we set the similar setup for domino arrangement where the desk height is 77 cm, a projector (EPSON dreamio EF-100, 2000 lumen), a Kinect V2 depth camera, and a pen tablet (Wacom Intous CTL-6100WL). The projector was placed 89 cm above the desk, while the height of camera was 80cm. 
The size of the lunch box was 20 cm $\times$ 13.5 cm $\times$ 4 cm. Because the light reflection may influence the captured infrared image, we placed a kitchen paper inside the box. We adjusted the canvas size of drawing interface to match the size ratio of the lunch box to be used. We set the canvas size to be 600px $\times$ 405px as shown in Figure \ref{fig:drawing}. 

\subsubsection{Projection Guidance}
The color selection is decided according to the colors of the prepared food ingredients. We investigated the common food ingredients which are often used in character bento production and determined the color of the ingredients correlated to the sketch. The ingredients used in the prototype system were broccoli, fried egg, crab stick, and sausage. Therefore, the drawing interface allows the use of green, yellow, orange, and pink colors according to the above ingredients.  

We use the depth sensor to detect the positions of food ingredients placed by the user. The depth map of the workspace without any ingredients is used as the environment map. In the comparison with the depth map with food ingredients, the environment map is excluded for noisy data. In our implementation, we decided the depth difference of 8.0 mm as the threshold value to detect the placement of food ingredients. If 70\% of the projected color area has been filled with food ingredients, the arrangement is considered to be correct with the projected green color. On the contrary, the red color will be projected when the ingredients are placed outside of the target area. The arrangement is considered to be completed if the area of red color is less than 20\% of the non-target area. The arrangement area then switches to the next step with a different color. If the subtasks of all colors are completed, a white color is projected.

\subsubsection{User Study} 
We conducted a comparison study to verify the effectiveness of the arrangement support with the prototype system. 12 participants were recruited in two groups: Group A with the proposed system; Group B without the proposed system. All the participants were asked to arrange a lunch box. We evaluated the time costs and the levels of completion with a questionnaire in five-point scale (``did you complete your lunch box the way you wanted?"). For Group B, the participants created bento by drawing the blueprint on paper as a reference. For Group A, the participants used the guidance of the proposed system. All the participants were asked to fill a questionnaire about the sketching interface and the arrangement support system after the task. In our questionnaire survey, we designed three questions items with 5-Point Likert scale (1 for strongly disagree, 5 for strongly agree). The questions include: Q1: Did you follow the guidance?; Q2: Easy to observe the guidance; Q3: Easy to understand the guidance.

\section{Results}
In this section, the task instruction and evaluation results of two case studies developed in SketchMeHow framework are discussed, and we explore the further potential application with the proposed framework. 

\subsection{Domino Arrangement}
\begin{figure}[t]
\centering
\includegraphics[width=1.0\linewidth]{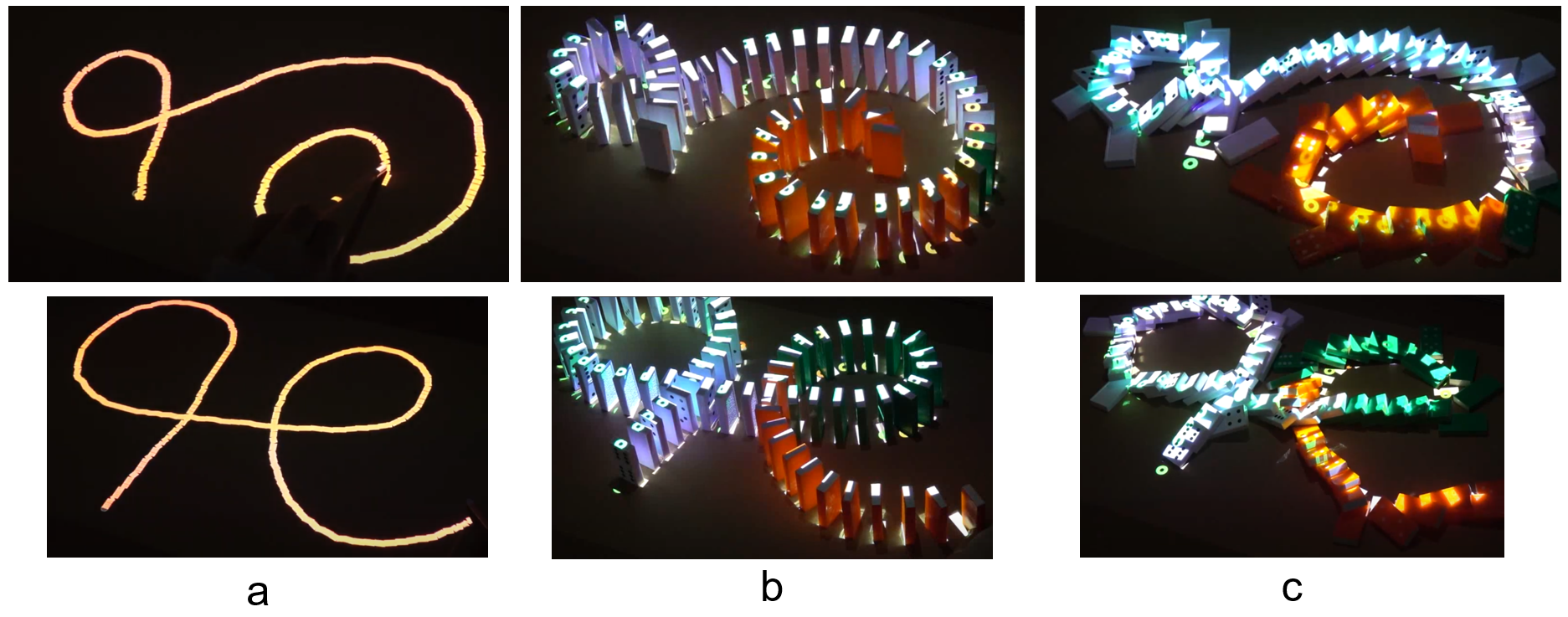}
\caption{The collapsed results (c) of domino arrangement (b) under user sketches (a).}
\label{fig:dominoresult2}
\end{figure}

\begin{figure}
\centering
\includegraphics[width=0.9\linewidth]{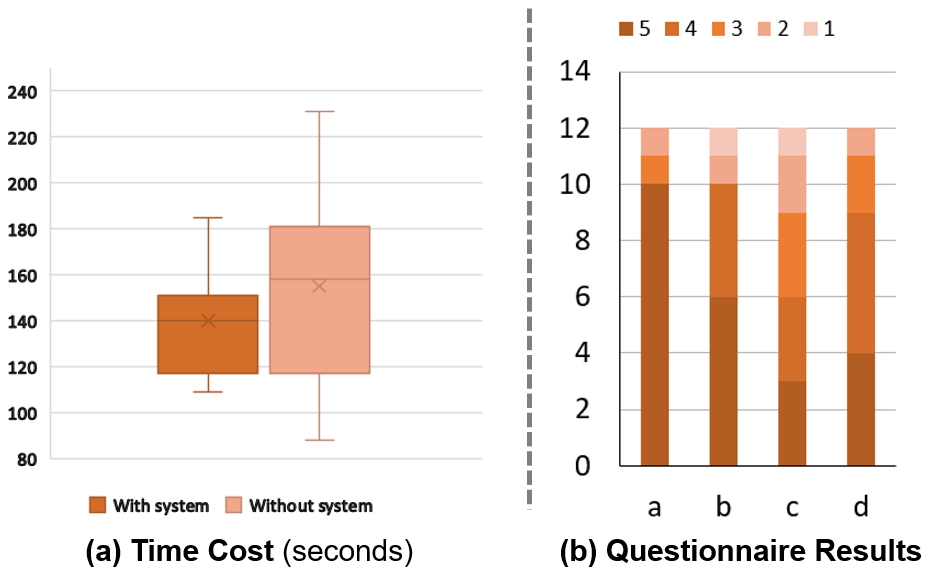}
\caption{Arrangement time cost (a) and the results of questionnaire (b).}
\label{fig:dominoresult}
\end{figure}

Figure \ref{fig:dominoresult2} shows the final arrangements of the domino matched the participants' sketches well. The operational time cost of with/without the domino arrangement system is shown in Figure \ref{fig:dominoresult}(a) . The average time cost using our system was 140 seconds (min 109 s, max 185 s). The average time cost without the system was 155 seconds (min 88 s, max 231 s). The variation and average of the processing time costs were reduced using the proposed system. We counted the number of participants who accidentally knocked down the domino blocks during arrangement which had 4 using the proposed system, and 8 without the system. Therefore, it is verified that SketchMeHow can not only shorten the arrangement time cost but also provide stability. 

Figure \ref{fig:dominoresult}(b) shows the results of the questionnaire in 5-Point Likert scale with the prototype system. It is verified that the proposed system has higher evaluation than without the system except for question c). In the case of the sense of accomplishment, the visual guidance that asked the participants to follow the projected positions may make participants feel boring. This issue can be improved by providing users more freedom of the modification in stroke drawing and domino blocks arrangement.

\subsection{Bento Production}
\begin{figure}
\centering
\includegraphics[width=1.0\linewidth]{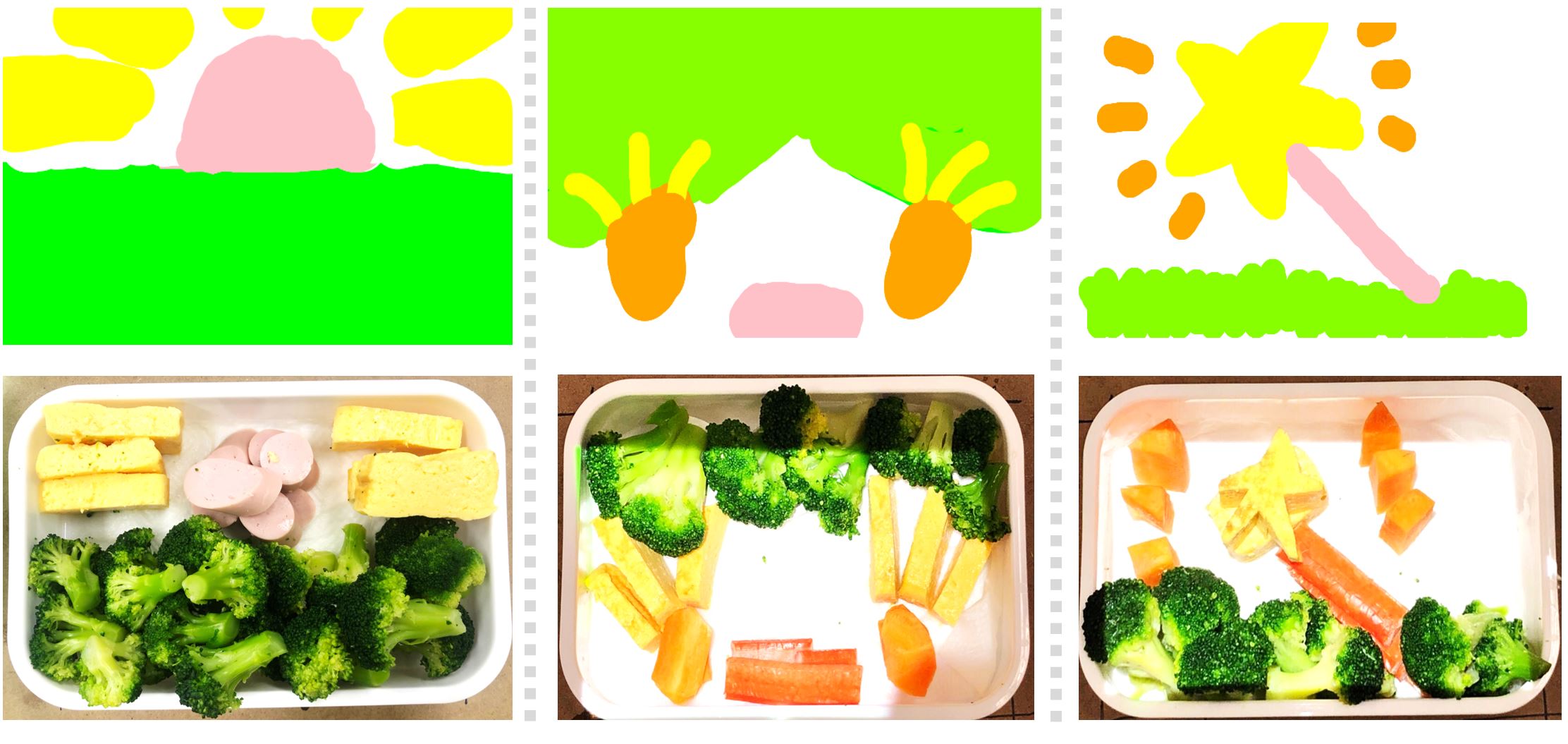}
\caption{Users sketches and the arranged bento box using the proposed system. }
\label{fig:bentoresult}
\end{figure}

\begin{figure}[htb]
\centering
\includegraphics[width=1.0\linewidth]{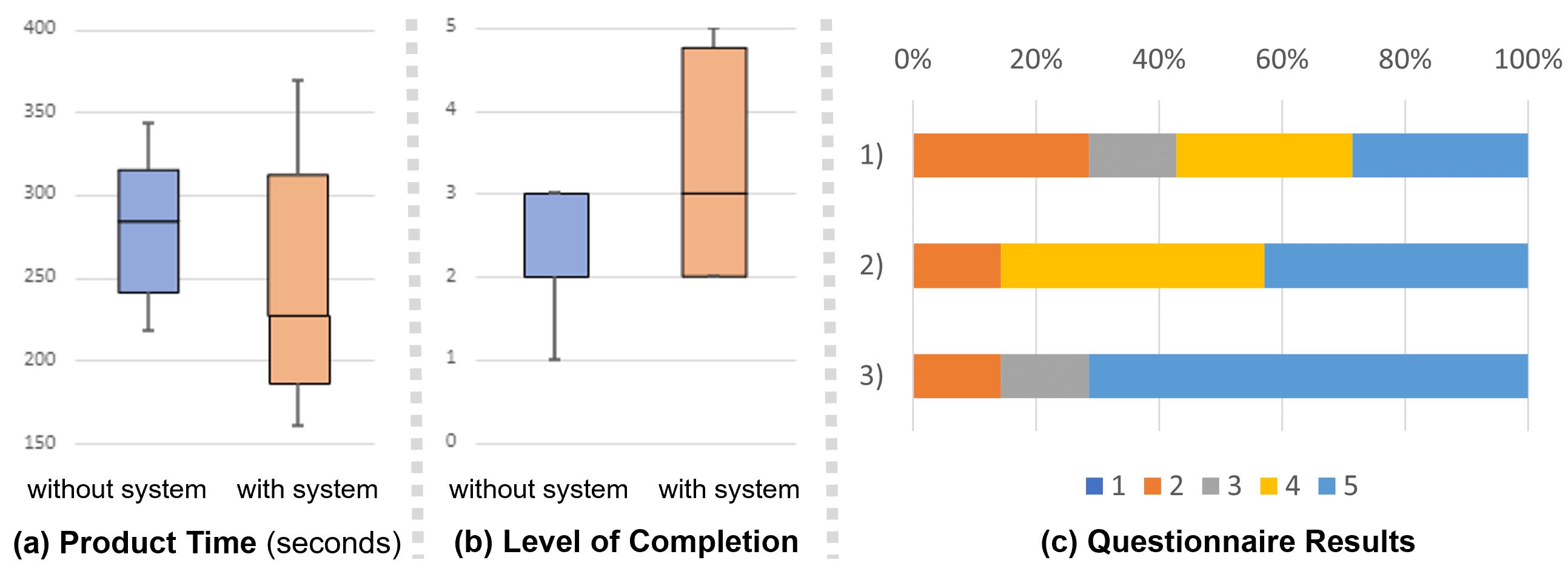}
\caption{From left to right: the production time costs (a), level of completion (b), and the questionnaire results (c).}
\label{fig:bentoscore}
\end{figure}

Figure \ref{fig:bentoresult} shows examples of bento produced in the proposed SketchMeHow framework. Figure \ref{fig:bentoscore} shows the comparison results of production time cost of with/without the proposed system and level of completion (5 for the highest degree of completion). The average production time cost of Group B is 280.2 seconds, while that of Group A with the proposed system was 250.6 seconds. We found that the production time cost decreased with the proposed system. For the level of completion, the average score for Group B was 2.4, while Group A was 3.3. Group B achieved lower scores because the participants had difficulty to arrange ingredients without any guidance. The participant spent more time in production because the participants may hesitate to adjust the positions of ingredients to match the sketch.

Figure \ref{fig:bentoresult}(c) shows The results of questionnaire that the scores of Q2 (4.1) and Q3 (4.3) were relatively higher than Q1 (3.6), which confirmed most of the participants were satisfied with the system guidance and the prototype system. Scores for Q1 was above 3.0, so the participants can follow the guidance well. Meanwhile, we found that the size of the lunch box is small and lack of ingredient choice in the study. The user may fail to fit the ingredients to the prompted position due to the inaccuracy between the food size and the prompted area. 

\subsection{Discussion}

\subsubsection{Cooperative Instruction System} In contrast to the instruction systems with head-mounted displays and tablets \cite{instruction2016,instruction2021}, SketchMeHow can support not only individual use and also cooperative working styles. It is possible for multiple users to collaborate in sketch drawing and task completion.

\subsubsection{Real World Applications} Not limited to the prototype systems developed in this work, we believe various real world applications may benefit from the proposed SketchMeHow framework, such as assembly tasks of hand craft and furniture construction, and fabrication work of hand-made objects and arts.  

\subsubsection{Large Scale System} In our case study, the desk-based tasks were adopted for framework verification. However, it is also feasible to support large scale instruction system, such as architecture-scale fabrication. For large-scale systems, the volumetric projection calibration may be required for accurate visual guidance, such as layered projection mapping technique \cite{xie2019balloonfab}.

\section{Conclusion}
In this work, we proposed a general sketch-based task instruction framework, SketchMeHow with the users' hand-drawn sketches as system inputs. The framework consists of the drawing interfaces in graphical and physical styles, the computation platform for task analysis, and the projection guidance with projector-camera systems. To verify the proposed SketchMeHow framework, we conducted two case studies of domino arrangement and bento production. In terms of the user studies in two prototype systems, it is verified that SketchMeHow framework can help novice users in task instruction in low computation cost and high completion levels. The participants were satisfied with the usage and visual guidance provided by the prototype systems.     

Based on the free comments in our case studies, the participants reported that the accuracy of recognition should be improved. In future, we plan to adopt depth sensor with higher resolution of depth images than the Kinect, such as the RealSense D455 depth camera. We believe that a higher resolution depth sensor can also reduce the noise data in prototype implementation. Under the SketchMeHow framework, the instruction system with user sketches can improve both the work efficiency and user satisfaction of our daily activities.  

\section*{Acknowledgement}
The authors thank the anonymous reviewers for their valuable comments. We greatly thank Yuki Mishima, Yamato Igarashi, Ryoma Miyauchi, Masahiro Okawa, Haruka Kanayama, Shogo Yoshida, Shogo Okada for idea discussion and experiment design. We would thank Ze Wang for sharing the codes of drawer interface. This project has been partially funded by JAIST Research Grant, Hayao Nakayama Foundation for Science \& Technology and Culture Grant A and JSPS KAKENHI grant JP20K19845, Japan.

\bibliographystyle{splncs04}
\bibliography{reference}
\end{document}